\documentclass[twocolumn,showpacs,prl,amsmath,amssymb,superscriptaddress]{revtex4}
\usepackage{graphicx}
\usepackage{bm}

\DeclareMathOperator{\Max}{Max}

\DeclareMathOperator{\Arctan}{arctan}

\begin{document}
\title{Ohmic and step noise from a single trapping center hybridized with a Fermi sea}
\author{Rogerio de Sousa}
\affiliation{Department of Chemistry
  and Pitzer Center for Theoretical Chemistry, University of
  California, Berkeley, California 94720-1460, USA} \author{K. Birgitta Whaley}
\affiliation{Department of Chemistry and Pitzer Center for Theoretical
  Chemistry, University of California, Berkeley, California
  94720-1460, USA}
\author{Frank K. Wilhelm} \affiliation{Department Physik, CeNS, and ASC,
  Ludwig-Maximilians-Universit\"at, Theresienstr. 37, D-80333
  M\"unchen, Germany} \author{Jan von Delft} \affiliation{Department
  Physik, CeNS, and ASC, Ludwig-Maximilians-Universit\"at, Theresienstr. 37,
  D-80333 M\"unchen, Germany} \date{\today}
\begin{abstract}
  We show that single electron tunneling devices such as the
  Cooper-pair box or double quantum dot can be sensitive to the
  zero-point fluctuation of a single trapping center hybridized with a
  Fermi sea.  If the trap energy level is close to the Fermi sea and
  has line-width $\gamma > k_BT$, its noise spectrum has an Ohmic
  Johnson-Nyquist form, whereas for $\gamma < k_B T$ the noise has a
  Lorentzian form expected from the semiclassical limit.  Trap levels
  above the Fermi level are shown to lead to steps in the noise
  spectrum that can be used to probe their energetics, allowing the
  identification of individual trapping centers coupled to the device.
\end{abstract}
\maketitle

Currently vigorous efforts are underway to achieve robust coherent
control over artificial two level systems (TLS) based on single charge
tunneling in superconducting \cite{nakamura99} or quantized
semiconductor \cite{hayashi03} islands.  While the main motivation is
to build a prototype for future quantum information technologies, the
associated need for unprecedented isolation from external noise also
make TLS structures sensitive probes of fundamental fluctuations in
the solid state \cite{schoelkopf02}. An interesting example is the
recent charge echo experiment by Nakamura and collaborators
\cite{nakamura02}, for which it has been claimed in \cite{galperin03}
that a \emph{single} background charge trapping center, henceforth
called ``local level'', is responsible for most of the coherence
decay, rather than 1/f noise originating from several charge traps
\cite{paladino02,nakamura02}. Although 1/f noise is usually
predominant in macroscopic samples, the signature of its individual
fluctuators is often identified in sensitive mesoscopic devices
\cite{savo87,fujisawa00}. The standard phenomenological description of
Random Telegraph Noise (RTN) of a fluctuating variable $\xi$ arising
from a single local level assumes a noise spectrum
given by \cite{kogan96,galperin03,itakura03}
\begin{equation}
\tilde{S}_\xi(\omega)=\frac{1}{2\pi}\int \textrm{e}^{i\omega t}\left\langle\xi(t)\xi(0)
\right\rangle dt
=\frac{\tilde{S}(0)}{(\omega\tau_c)^{2}+1},
\label{phrtn}
\end{equation}
where the trap fluctuation time scale $\tau_c$ can in principle be
computed microscopically.  The phenomenological Eq.~(\ref{phrtn})
should be contrasted with the universal Johnson-Nyquist (J-N) voltage
noise of a circuit with impedance $Z$ arising from particle-hole
excitations in a conductor, $\tilde{S}_{V}(\omega)= {\rm Re}(Z)
\hbar\omega\coth{(\hbar\omega/2k_BT)}$ 
\cite{callen51,weiss93}.  At high frequencies ($\hbar\omega\gg k_BT$)
the linear dispersion (Ohmic, or more generally $f$-noise) regime
resulting from J-N noise was measured long ago with the help of the
Josephson effect \cite{koch81}.  Here we show how a microscopic model
that in the semiclassical limit, whose regime of applicability will be
clarified later, leads to RTN [Eq.~(\ref{phrtn})], crosses over to
$f$-noise at low temperatures. Monitoring the excited state population
of a single charge tunneling device allows the detection of the
cross-over from semiclassical to quantum fluctuations.

We focus on the noise generated by a local trapping center (TC), e.g.
a dangling bond, located in the dielectric barrier close to one of the
electrodes \cite{paladino02,bauernschmitt93}.  When the trap energy
level $\tilde{\epsilon}$ is close to the Fermi level and the
temperature is lower than its line width $\gamma\sim\hbar/\tau_c$, we
find that the zero-point fluctuation of the trap generated by coupling
to the Fermi sea becomes evident through the appearance of
\emph{universal} linear dispersion in the noise spectrum which
reflects the Ohmic spectrum of electron-hole excitations in the
underlying Fermi sea.  At high temperatures we recover
Eq.~(\ref{phrtn}).  Furthermore, we show that trap levels above the
Fermi level ($\tilde{\epsilon}>\epsilon_F$) reveal themselves as steps
in the noise spectrum, so that multiple levels may give rise to a
staircase spectrum.  We discuss the relevance of these results for
recent measurements of noise by fast single-shot detection of the
excited state population of a Cooper-pair box \cite{astafiev04}.

{\it The model.} The Hamiltonian describing electrostatic coupling between a single
electron tunneling device in the TLS regime and a TC is
given by
\begin{eqnarray}
{\cal H}&=&\frac{1}{2}\delta\! E_C\sigma_z + \frac{1}{2}E_J \sigma_x + 
\left(\epsilon_d+\lambda
\sigma_z\right) d^\dag d
+\sum_k \epsilon_k c^{\dag}_{k}c_{k}
\nonumber\\
&&+\sum_k \left(V_{dk}+V'_{dk}\sigma_z\right) \left( d^\dag c_k
+ c^{\dag}_{k}d\right),
\label{htotal}
\end{eqnarray}
where $\sigma_i$ are Pauli operators acting on the states
$\mid\downarrow\rangle$ and $\mid\uparrow\rangle$, corresponding
respectively to zero and one excess Cooper-pair in the superconducting
island \cite{nakamura99,nakamura02}. The model also applies to an
electron localized in the left or right dot of a double quantum dot
structure \cite{hayashi03}.  The last two terms of Eq.~(\ref{htotal})
describe the tunneling of an electron in one of the metallic gates
controlling the TLS to a TC located in the dielectric interface
\cite{paladino02,bauernschmitt93}.  $d^\dag$ is a creation operator
for an electron in the TC, whose energy can assume the values
$\epsilon_d\pm\lambda$ depending on the state of the TLS. The matrix
element $V_{dk}\pm V'_{dk}$ models the tunneling amplitude between TC
and an electron in the metallic gate ($c_{k}^{\dag}$) with energy
$\epsilon_k$ ($c^{\dag}_k$ anti-commutes with $d^{\dag}$).  The
coupling between TLS and TC arises due to the modulation of TC
parameters modeled by the electrostatic coupling constants $\lambda$
and $V'_{dk}$. Previous work \cite{paladino02} did not contain the
$V'_{dk}$ term in Eq.~(\ref{htotal}), which arises due to the
sensitivity of the TC wave function to the TLS state (see estimates
below).  For simplicity we study a model of spinless electrons, as is
appropriate if TC double occupation is impossible due to Coulomb
repulsion in Eq.~(\ref{htotal}).  The last three terms of
Eq.~(\ref{htotal}) comprise the spinless Fano-Anderson Hamiltonian
\cite{fano_anderson}, which can be diagonalized exactly in the case
$\lambda=V'_{dk}=0$.  This is achieved by the transformation
\begin{equation}
d^\dag=\sum_k \nu_k \alpha^{\dag}_{k},\;\; c_{k}^{\dag}=\sum_{k'}
\eta_{k,k'}\alpha^{\dag}_{k'},
\label{eq:bogoliubov}
\end{equation}
where $\alpha_k$ are dressed electron operators.  In the thermodynamic
limit $\sum_k\rightarrow \int d\epsilon g(\epsilon)$, with the bare
electron density of states
$g(\epsilon)=\sum_k\delta_{\epsilon,\epsilon_k}/\delta\!\epsilon_k
=(3N/2\epsilon_F)\sqrt{\epsilon/\epsilon_F}$.  Here $\delta\!\epsilon_k$
is the mean energy spacing between non-degenerate levels of the
electron gas and $\delta_{\epsilon,\epsilon_k}$ is the Kronecker
delta.  The continuous description of $\nu_k$ is then given by the
spectral function $A(\epsilon)= \sum_{k}
\frac{\nu_k^2}{\delta\!\epsilon_k} \delta_{\epsilon,\epsilon_k},$ which
is interpreted as the mean density in energy of the admixture of the
TC with the electron bath.  The energy density for the electrons
which do not participate in the admixture with the TC is given by
$g'(\epsilon)=\sum_{k,k'}\frac{\left|\eta_{k',k}\right|^2}{\delta\!\epsilon_{k}}
\delta_{\epsilon,\epsilon_{k}}$. Below we show that the response of
the TLS to a TC depends only on the densities $A(\epsilon)$
and $g'(\epsilon)$. 
The exact solution for $A(\epsilon)$ in the
thermodynamic limit has the Lorentzian form \cite{fano_anderson}
\begin{equation}
A(\epsilon)= \frac{\gamma/\pi}{\left(\epsilon - \tilde{\epsilon}\right)^2+\gamma^2},
\label{alor}
\end{equation}
normalized to unity by virtue of the commutation relation
$\{d,d^{\dag}\}=1$.  
The energy densities are related by
\begin{equation}
g'(\epsilon)=g(\epsilon)+\frac{d\ln{g(\epsilon)}}{d\epsilon}
\left(\epsilon-\tilde{\epsilon}\right)A(\epsilon).
\end{equation}
Here the dressed trap energy $\tilde{\epsilon}$
differs from $\epsilon_d$ to second order in $V_{dk}$. The line-width
$\gamma=\pi V_{\tilde{\epsilon}}^2g(\tilde{\epsilon})$ is essentially
the Fermi Golden rule for the decay of a localized state (TC) into the
continuum (Fermi sea), with $V^{2}_{\tilde{\epsilon}}\equiv \langle
V^{2}_{dk}\rangle_{\epsilon_k=\tilde{\epsilon}}$.  
Finally we perform a rotation on the Pauli
operators to get the transformed Eq.~(\ref{htotal}),
\begin{eqnarray}
{\cal H}&=&\frac{1}{2}\hbar\omega \sigma_z
+\left(\sin{\theta}\sigma_x+
\cos{\theta}\sigma_z\right)\sum_{k,k'}
\biggl[
\lambda\nu_k\nu_{k'}\alpha_{k}^{\dag}\alpha_{k'}\nonumber\\
&&+V'_{dk'}\sum_{k''}(\nu_k \eta_{k',k''}+\nu_{k''}
\eta_{k',k})\alpha_{k}^{\dag}\alpha_{k''}\biggl]\nonumber\\
&&+\sum_k\epsilon_k \alpha^{\dag}_k\alpha_k,
\label{htrans}
\end{eqnarray}
with $\tan{\theta}=E_J/\delta \!E_C$ and
$\hbar\omega=\sqrt{E_{J}^2+\delta\!E_C^2}$. 
Eq.~(\ref{htrans}) has the structure of a TLS coupling to
electron-hole excitations in a new, effective Fermi sea, characterized
by coupling constants $\lambda\nu_k\nu_{k^\prime}$.

\begin{figure}
\includegraphics[width=3.4in]{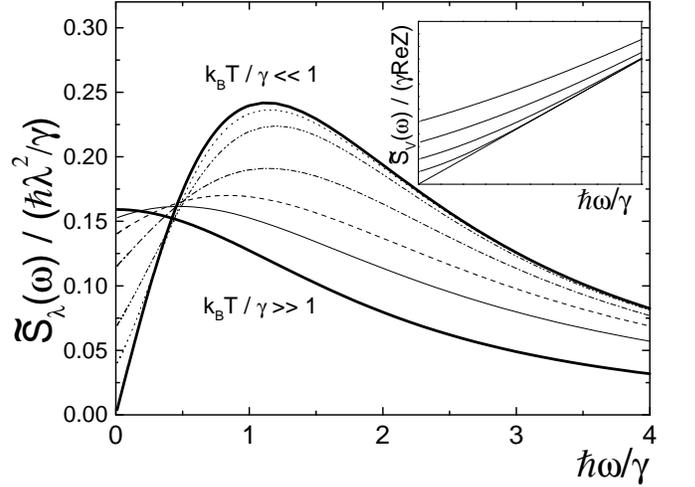}
\caption{Noise spectral density 
  describing energy transfer from a TLS to a single trap level close
  to the Fermi sea for $k_BT/\gamma=0,0.1,0.2,0.5,1,2,\infty$.  At
  high temperatures we have the standard random telegraph noise
  spectrum.  As the temperature is lowered below the trapping center
  line-width $\gamma$, Ohmic dispersion can be observed in the low
  frequency regime.  This behavior is very similar to the
  Johnson-Nyquist spectrum (inset).}\label{fig1}
\end{figure}

{\it Ohmic J-N noise.} The decay rate for a TLS initially prepared in the excited state
$\mid\uparrow\rangle$ gives a measure of the noise spectral density at
frequency $\omega$ through the golden-rule result
\cite{schoelkopf02}
\begin{equation}
\frac{1}{T_1}=\frac{\pi\sin^2{\theta}}{2\hbar^2}
\left[
\tilde{S}_{\lambda}(\omega)+\tilde{S}_{V'}(\omega)
+\tilde{S}_{\lambda}(-\omega)+\tilde{S}_{V'}(-\omega)
\right].
\label{t1}
\end{equation}
Note that the weak coupling between the detector (TLS) and the noise
source (TC) implies the TC remains in thermal equilibrium with the
Fermi sea \cite{weiss93}.  Here $\tilde{S}$ are Fourier transforms of
the time-ordered correlation functions as in Eq.~(\ref{phrtn}).
$\tilde{S}_{\lambda}$ occurs due to the charging and discharging of
the TC, where $\bar{n}=\langle d^{\dag}d\rangle=\int
A(\epsilon)f(\epsilon)d\epsilon$ is the average TC occupation, and
$f(\epsilon)=1/[1+\textrm{e}^{(\epsilon-\epsilon_F)/k_BT}]$ the Fermi
function.  The noise densities are calculated by
Eq.~(\ref{eq:bogoliubov}) and (\ref{htrans}),
\begin{eqnarray}
\tilde{S}_{\lambda}(\omega)&=&\frac{4\hbar\lambda^2}{2\pi} \int dt \textrm{e}^{i\omega t}
\left\langle \left[d^\dagger(t)d(t)-\bar{n}\right]
\left[d^\dagger(0)d(0)
-\bar{n}\right]\right\rangle_{\rm FS}\nonumber\\
&=&4\hbar\lambda^2\int d\epsilon A(\epsilon)A(\epsilon-\hbar\omega)
f(\epsilon-\hbar\omega)[1-f(\epsilon)].
\label{slambda}
\end{eqnarray}
Here, the $d$, $d^\dagger$ are taken in the interaction representation and
$\left\langle\cdot\right\rangle_{\rm FS}$ denotes averaging over the free Fermi sea.
A similar expression applies for $\tilde{S}_{V'}$, with the
operator $d^{\dag} d$ in Eq.~(\ref{slambda}) substituted by
$\sum_k V'_{dk}d^{\dag}c_k +\rm{h.c.}$.  Because we assume the TC is
in thermal equilibrium, Eq.~(\ref{slambda}) satisfies the detailed
balance condition $\tilde{S}(-\omega)=\exp{(-\hbar\omega/k_B
  T)}\tilde{S}(\omega)$.  Moreover, the total noise $\int
\tilde{S}_{\lambda}(\omega)d\omega$ is equal to the mean square
fluctuation $S_{\lambda}(0)=4\lambda^2(\bar{n}-\bar{n}^2)$, which is
appreciable only if $\bar{n}$ is not close to zero or one.  This
happens only if $|\tilde{\epsilon}-\epsilon_F|\lesssim \Max{\{k_B
  T,\gamma\}}$, so Eq.~(\ref{slambda}) is only of appreciable size if
$\tilde{\epsilon}$ is close enough to the Fermi energy. For
$\tilde{\epsilon}=\epsilon_F$ we have $\bar{n}=1/2$ and
$S_{\lambda}(0)=\lambda^2$ is maximum.  Consider the noise spectrum at
high temperature, $k_BT\gg \gamma$.  Because the integral in
Eq.~(\ref{slambda}) is only appreciable within the range of
$A(\epsilon)$, the Fermi functions can be approximated by $1/2$ and
the resulting integral easily integrated by the residue method.
Therefore the high temperature noise is determined solely by the pole
structure of $A(\epsilon)$. Using Eq.~(\ref{alor}) we recognize the
spectrum for random telegraph noise [Eq.~(\ref{phrtn})] with
correlation time $\tau_c=\hbar/(2\gamma)$.

Now consider the opposite limit of $k_BT\ll \gamma$.  For high
frequencies ($\hbar\omega\gg
\Max{\{|\tilde{\epsilon}-\epsilon_F|,\gamma\}}$) we have the
asymptotic behavior $\tilde{S}_{\lambda}(\omega)\approx
A(\epsilon_F-\hbar\omega)(1-\bar{n})\sim 1/\omega^2$.  For positive low
frequencies ($0\leq\hbar\omega\ll \gamma$) but arbitrary
$\hbar\omega/k_BT$, we may approximate Eq.~(\ref{slambda}) by
\begin{eqnarray}
\tilde{S}_{\lambda}(\omega)&\approx&
4\hbar\lambda^2
A\left(\epsilon_F\right)^2
\int d\epsilon f(\epsilon-\hbar\omega)[1-f(\epsilon)]\nonumber\\
&=&4\hbar^2\lambda^2
A\left(\epsilon_F\right)^2\omega
\left[n(\omega)+1\right],
\label{lambdaOhmic}
\end{eqnarray}
where $n(\omega)=1/[\textrm{e}^{\hbar\omega/k_BT}-1]$ is the Bose function.
Using detailed balance we get 
\begin{equation}
\frac{\tilde{S}_{\lambda}(\omega)+\tilde{S}_{\lambda}(-\omega)}{2}
\approx 2\hbar^2\lambda^2 A\left(\epsilon_F\right)^2\omega
\coth{\left(\frac{\hbar\omega}{2k_BT}\right)},
\end{equation}
which makes evident the universal behavior of low frequency noise
arising due to a TC interacting with a Fermi sea, i.e., $\tilde{S}$
has the Johnson-Nyquist form for small $\omega$, independent of the
functional form of $A(\epsilon)$, provided that this is analytic at
$\epsilon_F$.  The transition between Eqs.~(\ref{phrtn}) and
(\ref{lambdaOhmic}) is shown in Fig.~\ref{fig1}.  This behavior is a
signature of the crossover from zero-point quantum to classical
fluctuations for the electron-hole excitations whose energy $\hbar\omega$
lies within the hybridization bandwidth $\gamma$.  In other words, the
TC acts as a filter of bandwidth $\gamma$, allowing only electron-hole
excitations of energy $|\hbar\omega|\lesssim \gamma$ to affect the TLS.
When $k_B T<\gamma$, the environment behaves effectively like a flat
band Fermi gas, with bosonic electron-hole pair excitations leading to
Johnson-Nyquist noise.  However, when $k_BT>\gamma$ the frequency of
fluctuations with $|\hbar\omega|\lesssim \gamma$ are always within the
semiclassical regime ($\hbar\omega<k_B T$), which leads to incoherent
hopping of the TC charge into and out of the Fermi sea, giving rise to
random telegraph noise.

\begin{figure}
\includegraphics[width=3.4in]{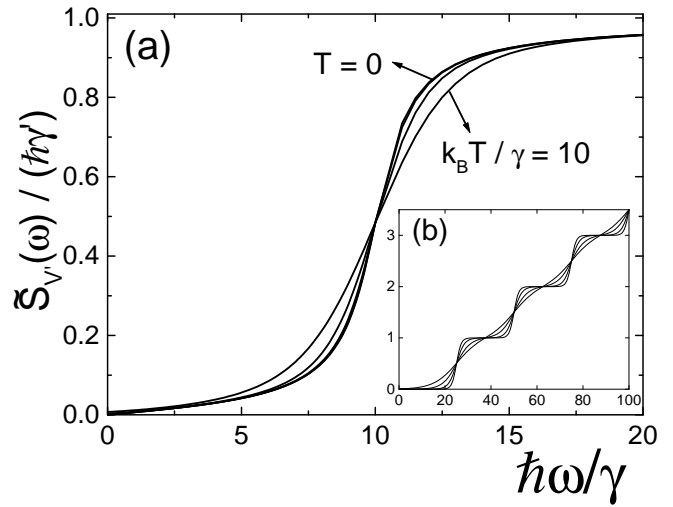}
\caption{(a) Noise spectrum from a single trapping center away from the
  Fermi sea. The step feature occurs at
  $\omega=|\tilde{\epsilon}-\epsilon_F|/\hbar$ with step-width given
  by $\Max{\{\gamma,k_BT\}}$.  (b) Staircase noise arising from
  several traps with energy levels equally spaced. Here as the
  temperature increases (with respect to the level separation), the
  noise appears to be linear in frequency.}
\label{fig2}
\end{figure}

{\it Staircase noise.} We now turn to the effects of the $V'_{dk}$
interaction.  In the original Eq.~(\ref{htotal}) this term changes the
hybridization coupling of the TC and Fermi sea, which transforms into
a TLS-dependent modification in the dressed trap energy
$\tilde{\epsilon}$. A calculation similar to Eq.~(\ref{slambda})
yields
\begin{eqnarray}
\tilde{S}_{V'}(\omega)&=&4\hbar\int d\epsilon \left[
V'^{2}_{\epsilon-\hbar\omega}g'(\epsilon-\hbar\omega)A(\epsilon)
+V'^{2}_{\epsilon}g'(\epsilon)
\right.\nonumber\\
&&\left.\times
A(\epsilon-\hbar\omega)\right]
f(\epsilon-\hbar\omega)[1-f(\epsilon)].
\label{svpc}
\end{eqnarray}
In contrast to the $\lambda$-processes, this rate is finite for
trap energies $\tilde{\epsilon}$ far from the Fermi level,
i.e. $|\tilde{\epsilon}-\epsilon_F|\gg \Max{\{\gamma,k_BT\}}$: 
The first term dominates if
$\tilde{\epsilon}>\epsilon_F$, while the second dominates if
$\tilde{\epsilon}<\epsilon_F$.  For realistic parameters
$g'(\epsilon)\approx g(\epsilon_F)$ (since $\hbar\omega\ll \epsilon_F$), we
may approximate Eq.~(\ref{svpc}) by
\begin{equation}
\tilde{S}_{V'}(\omega)\approx
\hbar\gamma' \int d\epsilon
\left[A(\epsilon)+A(\epsilon-\hbar\omega)\right]
f(\epsilon-\hbar\omega)[1-f(\epsilon)],
\label{svpa}
\end{equation}
where $\gamma'=\pi V'^{2}_{\epsilon_F}  g(\epsilon_F)$ is a line-width
associated with $V'_{dk}$. $\tilde{S}_{V'}(\omega)$ has a step at
$\omega=|\tilde{\epsilon}-\epsilon_F|/\hbar$, with step-width given
approximately by $\Max{\{\gamma,k_B T\}}$. For low temperatures ($k_B
T\ll \gamma$) Eq.~(\ref{svpa}) is controlled through the phases of electrons
scattering off the local level and reads
\begin{eqnarray}
\tilde{S}_{V'}(|\omega|)&=&\frac{\hbar\gamma'}{\pi}\left[
\Arctan{\left(\frac{\tilde{\epsilon}-\epsilon_F+|\hbar\omega|}{\gamma}\right)}\right.\nonumber\\
&&\left.-\Arctan{\left(\frac{\tilde{\epsilon}-\epsilon_F-|\hbar\omega|}{\gamma}\right)}
\right].
\end{eqnarray}
At high frequencies ($\omega\gg |\tilde{\epsilon}-\epsilon_F|/\hbar$)
the noise spectrum saturates at $\simeq \hbar \gamma'$ for all
temperatures, as long as $\omega$ is less than a cut-off frequency,
arising either due to a cut-off in the matrix element $V^\prime$ or
the bandwidth.

It is straightforward to generalize Eq.~(\ref{htotal}) to the case of
several TC's tunneling to different gate electrodes
\cite{paladino02,bauernschmitt93}, each with energy
$\tilde{\epsilon}_i$ and line-width $\gamma_i$. If more than one TC is
located in the same gate, their correlation energy $E_{ij}$ (arising
due to Coulomb interaction or to inter-trap tunneling) can be
neglected provided $E_{ij} \ll
|\tilde{\epsilon}_i-\tilde{\epsilon}_j|$. In this case $V'$-noise
allows spectroscopy of the energy levels $\tilde{\epsilon}_i$ with
efficiency proportional to $\gamma'_i$. The resulting spectrum will be
a staircase with each step located at $\tilde{\epsilon}_i$ [See
Fig.~\ref{fig2}(b)]. This can be used for experimental determination
of the number of trap levels within a specified range from the
island. If the separation between levels is approximately equal, and
the experiment does not have enough resolution to resolve the steps,
the noise spectrum will look Ohmic.  In contrast to the previous case
of just $V$ coupling, approximate Ohmic behavior due to the
$V^\prime$-couplings now arises even when the high temperature
condition is satisfied ($k_BT\gg \gamma_i$).

{\it Estimated parameters.} To obtain order of magnitude estimates of
$\lambda$, $\gamma$, and $\gamma'$, consider a hydrogenic model for
the trapping center wave function, $\Psi_d(r)=\exp{(-r/a_B)}/\sqrt{\pi
  a_B^3}$, with Bohr radius $a_B\sim 15$~\AA.  The line-width $\gamma$
is calculated from the overlap integral with conduction electrons in
the gate electrode, leading to $\gamma \approx 96\pi^2 \epsilon_F n
a_B^3/(k_F a_B)^8$.  Using an electron density $n=10^{22}$~cm$^{-3}$
and $\epsilon_F\sim 1$~eV, we obtain $\gamma \sim 10$~$\mu$eV$\sim
k_B\times 100$~mK$\sim h\times 3$~GHz.  $\lambda$ is given by the
dipolar coupling between the TLS (dipole $e r_{12}$ with $r_{12}\sim
0.1$~$\mu$m) and the TC (dipole $e a_B$ due to the image charge
produced in the gate), leading to $\lambda\sim 0.03$~$\mu$eV when the
TLS-trap distance is $r\sim 0.5$~$\mu$m. This gives a contribution to
TLS relaxation of the order of $\frac{1}{T_1}\sim 10^5$~s$^{-1}$ at
low frequencies.  The step amplitude $\gamma'$ [Eq.~(\ref{svpa})]
derives from the distortion of the TC wave function (and consequently,
the overlap integral), by the electric field ${\cal E}$ produced by
the TLS, $V'_{dk}/V_{dk}\sim \delta\Psi_d/\Psi_d\sim e{\cal
  E}a_B/(e^2/a_B)$. If $r\sim 1$~$\mu$m this effect is very small,
$\gamma'/\gamma\sim a_{B}^4r_{12}^2/(\kappa^2 r^6)\sim 10^{-13}$ (here
the dielectric constant $\kappa= 10$).  However, TC's close to the TLS
($r\sim 0.01$~$\mu$m) are affected by a monopolar electric field, and
hence are given by $\gamma'/\gamma\sim (a_B/r)^4/\kappa^2\sim
10^{-5}$. This leads to a step amplitude of the order of
$\frac{1}{T_1}\sim 10^5 s^{-1}$ at high frequencies, which is high
enough for the number of TC's to be determined. Note that
$\gamma,\gamma',\lambda$ are strongly dependent on the TC Bohr radius.

The above results are valid in the weak-coupling regime, when
$\lambda\ll \gamma$ and $\gamma'\ll \epsilon_F$.  For $\lambda>\gamma$
the TLS dynamics acts back on the TC, for example
Eq.~(\ref{alor}) splits into two peaks at energies
$\tilde{\epsilon}_{\pm}\approx \tilde{\epsilon}\pm\lambda$
(zero-frequency noise was studied recently at the strong coupling
regime, see \cite{grishin04}). In this regime back-action effects are
important, and assuming the TC is in thermal equilibrium may not
be appropriate.

Remarkably, the ultraviolet cutoff of the Ohmic spectrum, as mentioned
above, is a power law, similar to the Drude cutoff of the standard
Ohmic model \cite{weiss93}. On the other hand, the high-temperature
regime of the Ohmic bath leads to a Lorentzian power spectrum as in
Eq.~(\ref{phrtn}).

{\it Relevance to experiments.}  Our theoretical result that even a
single TC hybridized with a Fermi sea can give rise to Ohmic noise
contrasts with previous studies that have assumed a distribution of
two-level fluctuators \cite{shnirman04}. The current study was
motivated by measurements of $f$-noise in a Cooper-pair box 
\cite{astafiev04} over the frequency range $\omega= 3-100$~GHz at
$T=50$~mK. Depending on the value of the TC Bohr radius we find that
at low temperatures we will have either Ohmic noise deriving from a TC
with energy close to the Fermi energy, or a step noise spectrum for a
TC well above or below the Fermi energy. The latter becomes a
staircase function for multiple traps. A more detailed experimental
analysis with greater resolution than ~\cite{astafiev04} is required
to differentiate between these two behaviors. One additional
possibility for measuring the predicted signature of individual
fluctuators is to vary the gate voltage configurations as in
Ref.~\cite{fujisawa00}.

{\it Conclusions.} We describe a scenario for $f$-noise that is based
on the interaction of a single electron tunneling device with one or
only a few trapping centers.  At the lowest temperatures
($T<\gamma/k_B\lesssim 100$~mK) we find that the low frequency noise
can be surprisingly different from the standard random telegraph
spectrum.  At high frequencies, we predict that multiple traps can
give rise to a stair-case signal arising from traps which are located
nearby the device but have energies far away from the Fermi level.
These could be a source of $f$-noise even at high temperatures, and
can be used to measure the number of traps within a certain distance
from the device.

We acknowledge support from NSF, DARPA SPINS, DFG, and ARDA. We thank
J. Clarke, I. Martin, E. Mucciolo, and Y. Pashkin for discussions.

%
%

\end{document}